\documentclass[aps,pre,reprint,onecolumn,notitlepage,superscriptaddress]{revtex4-2}
\usepackage{amssymb,amsmath,color,mathtools,subeqnarray,bm,mathtools,upgreek}
\usepackage[colorlinks=true,linkcolor=blue]{hyperref}
\usepackage[sfdefault=cmbr]{isomath}
\usepackage{graphicx}
\usepackage{epstopdf, epsfig}
\usepackage{xcolor}
\usepackage{subfig}
\usepackage{comment}

\def \t{\tensorsym}

\begin{document}

\title{Swimming efficiency in viscosity gradients}

\author{Jiahao Gong}%
\affiliation{%
 Department of Mathematics,\\
University of British Columbia, Vancouver, BC, V6T 1Z2, Canada
}%

\author{Vaseem A. Shaik}
\affiliation{
Department of Mechanical Engineering,\\
University of British Columbia, Vancouver, BC, V6T 1Z4, Canada
}%

\author{Gwynn J. Elfring}
 \email{gelfring@mech.ubc.ca}
\affiliation{%
 Department of Mathematics,\\
University of British Columbia, Vancouver, BC, V6T 1Z2, Canada
}%
\affiliation{
Department of Mechanical Engineering,\\
University of British Columbia, Vancouver, BC, V6T 1Z4, Canada
}%

\begin{abstract}
In this note, we study the effect of viscosity gradients on the energy dissipated by the motion of microswimmers and the associated efficiency of that motion. Using spheroidal squirmer model swimmers in weak linearly varying viscosity fields, we find that efficiency depends on whether they generate propulsion from the back (pushers) or the front (pullers). Pushers are faster and more efficient when moving down gradients but slower and less efficient moving up viscosity gradients, and the opposite is true for pullers. However, both pushers and pullers display negative viscotaxis, therefore pushers dynamically tend to the most efficient orientation while pullers the least. We also evaluate the effect of shape on power expenditure and efficiency when swimming in viscosity gradients and find that in general the change in both due to gradients monotonically decreases with increasing slenderness. This work shows how shape and gait play an important role in determining dynamics and efficiency in inhomogeneous environments, and demonstrating that both efficiency minimizing and maximizing stable dynamical states are possible.
\end{abstract}

\maketitle

\section{Introduction}
Swimming microorganisms are widely found in nature and important in diverse array of biological processes \citep{Vogel1996,Lauga2016,Wadhwa2022,Ishikawa2023i,Ishikawa2023ii}. An understanding of the dynamics of microswimmers and other active particles (biological or synthetic), in homogeneous Newtonian fluids is now reasonably well developed \citep{Brennen1977,Lauga2009,Yeomans2014,Elgeti2015,Ishikawa2024}. However, natural environments can often be quite complex and inhomogeneous \citep{Bechinger2016,Martinez-Calvo2023}. 

Inhomogeneity of fluid environments can arise due to spatial variations of different physical quantities, including light \citep{Jekely2009}, heat \citep{Bahat2003}, chemical concentration or nutrients \citep{Berg1972,Berg2004}, and can lead to directed motion, which is known as \textit{taxis}. Natural bodies of water, such as lakes, ponds, or oceans, often have gradients in temperature or salinity and these inhomogeneities can also lead to a stratification of the mechanical properties of the fluid such as density and viscosity \citep{Stocker2012}. In response to gradients in viscosity, bacteria such as \textit{Leptospira} and \textit{Spiraplasma} have been observed to perform positive viscotaxis in experiments \citep{Kaiser1975,Petrino1978,Daniels1980,Takabe2017}, while \textit{Escherichia coli} and \textit{Chlamydomonas reinhardtii} perform negative viscotaxis \citep{Sherman1982,Stehnach2021,Coppola2021}. A gradient in the viscosity of the intestinal mucosal barrier is thought to control the spatial organization of intestinal microbiota \citep{Swidsinski2007}.

While there is an obvious benefit associated with moving to environment where the energetic penalty of motion is lower, recent work has shown that such directed motion can arise as an immediate consequence of interaction with an environment with inhomogeneous mechanical properties. Inhomogeneity can break rotational symmetry and this leads naturally to reorientation in order to conserve angular momentum. In pioneering work, \citet{Liebchen2018} showed that active particles, modeled as linked spheres moving with a constant propulsive force, in weak viscosity gradients exhibited positive viscotaxis due to the mismatch of viscous drag on the spheres. Further investigations incorporated the impact of viscosity changes on propulsion using the spherical \citep{Datt2019,Shaik2021,Gong2023}, and spheroidal \citep{Gong2024} squirmer model, which simulates a swimming gait by a surface slip velocity. These studies found that the interaction between the particle's active slip and spatial variations of viscosity tends to dominate the dynamics and typically results in \textit{negative} viscotaxis, although the effect of the gradient decreases with increased slenderness.

Changes in translational and rotational dynamics that arise as a consequence of viscosity gradients are now relatively well understood, for both passive \citep{Kamal2023,Anand2024} and active particles \citep{Gong2024}, but here we focus on the power expended by microswimmers and any associated changes in efficiency that are caused by the inhomogeneous environment. Even though some microswimmers are capable of achieving higher speeds in viscosity gradients, we still need to quantify the resultant changes in power expenditure in order to determine whether such motion is in fact more efficient. We assess the efficiency of microswimmers using the Froude efficiency, which is defined as the ratio of the power expended to simply drag a body of the same shape to the power expended by propulsion \citep{Lighthill1952}. This efficiency has been broadly used to characterize different microswimmers in homogeneous Newtonian fluids \citep{Stone1996,Chattopadhyay2006,Ishimoto2014}, although more recently, other efficiency measures have been proposed \citep{Childress2012,Nasouri2021,Ider2023}, to guarantee a measure which can never exceed unity. 

Several studies have investigated the efficiency of microswimmers in homogeneous non-Newtonian environments. One study showed that pusher-type spherical squirmers always expend more power and swim less efficiently than puller-type squirmers in second order fluids \citep{Corato2015}. In fluids with shear-thinning properties, all types of two-mode spherical squirmers swim more slowly, expending less power with higher efficiency \citep{Nganguia2017}, but by introducing a third squirming mode it is possible to design a microswimmer which can swim faster and more efficiently in shear-thinning fluids. Specifically for mechanically inhomogeneous fluids, a swimming sheet was shown to expend more power and move with reduced efficiency in a viscosity stratified fluid in comparison to a homogeneous Newtonian fluid \citep{Dandekar2020}.

In this work, we aim to explore how viscosity gradients affect the swimming efficiency of spheroidal squirmers in order to understand the role of swimming gait and shape on power expended and efficiency in an inhomogenous fluid. In what follows we derive analytical formulas for the energy expended by these micoswimmers and the associated efficiency assuming weak constant gradients in the viscosity.
 
\section{Swimmers in viscosity gradients}
\subsection{The squirmer model}
Microswimmers are modelled as squirmers in this research. The spherical squirmer model is a classical hydrodynamic model for the motion of self propelling particles, such as protozoa or volvocine green algae, wherein the complex motions of ciliated surfaces are represented as a tangential slip velocity on the surface of a spherical body \citep{Lighthill1952,Blake1971}. The slip velocity is commonly written in terms of Legendre polynomials
\begin{equation}
    \boldsymbol{u}^s = - \sum_{n=1}^{\infty}  \frac{2 B_n}{n(n+1)} P_n'\left(\boldsymbol{p} \boldsymbol{\cdot} \boldsymbol{n} \right)\boldsymbol{p} \boldsymbol{\cdot}(\mathsf{\t I} - \boldsymbol{n} \boldsymbol{n} ),
    \label{eqn:spherical_us}
\end{equation}
where $\boldsymbol{p}$ is the swimming direction of the particle, $P_n$ is the Legendre polynomial of degree $n$ and $\boldsymbol{n}$ is a unit normal to the squirmer surface. The coefficients, $B_n$, often called squirming modes, can be related to Stokes flow singularity solutions \citep{Pak2014}. The swimming velocity of a particle is determined by the $B_1$ mode (here we assume $B_1 \geq 0$), while the $B_2$ mode sets the sets magnitude of the force dipole, the slowest decaying contribution to the far-field flow.

\begin{figure}
\centering
\includegraphics[scale = 0.5]{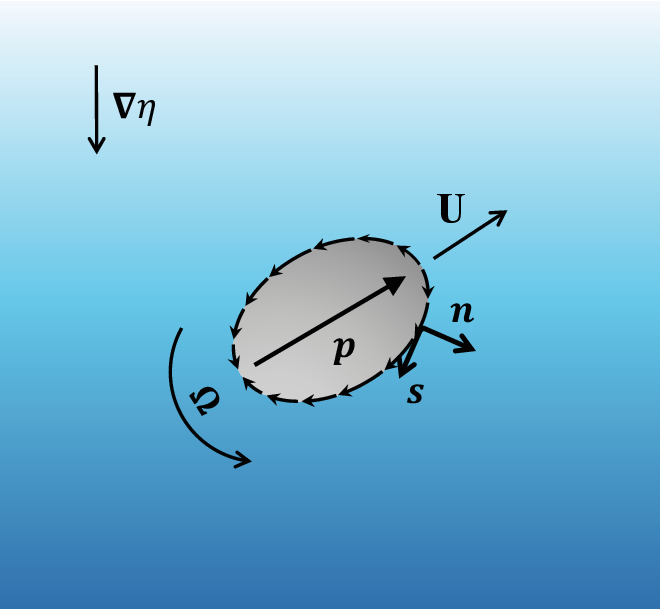} \hfill
\caption{Schematic of a spheroidal squirmer swimming in a constant viscosity gradient $\boldsymbol{\nabla} \eta$. $\boldsymbol{p}$ represents the swimming direction of the swimmer.}
\label{fig:schematic2}
\end{figure}

Since many organisms are elongated in shape, such as \textit{Paramecium caudatum} or \textit{Opalina}, the squirmer model was extended to accommodate spheroidal geometries by \citet{Keller1977} and the streamlines predicted by their spheroidal model indeed closely matched the experimental streak photographs of \textit{Paramecium caudatum}. The original model by \citet{Keller1977} only included the first squirming mode and latter studies added a force-dipole mode to better represent other types of microswimmers \citep{Gaffney2011,Ishimoto2014,Theers2016}. We use the two-mode spheroidal squirmer here to represent prolate spheroidal active swimmers \citep{Vangogh2022,Theers2016,Gong2024}. The slip velocity on the surface of a prolate spheroidal squirmer can be expressed as,
\begin{equation}
       \boldsymbol{u}^s  = -B_1 (\boldsymbol{s} \boldsymbol{\cdot} \boldsymbol{p}) \boldsymbol{s} - \frac{B_2}{a} (\boldsymbol{r} \boldsymbol{\cdot} \boldsymbol{p} ) (\boldsymbol{s} \boldsymbol{\cdot} \boldsymbol{p}) \boldsymbol{s},
       \label{eqn:spheroidal_us}
\end{equation}
where $\boldsymbol{r}$ is a vector from the center of the particle to a point on the surface while $\boldsymbol{s}$ is the unit tangent at that point (see Figure \ref{fig:schematic2}). This form is equivalent to the formula in \eqref{eqn:spherical_us} for a sphere with only two modes. Recent research points out the swimming speed and stresslet of such a squirmer are influenced by more than just the $B_1$ and $B_2$ modes \citep{Pohnl2020}; however, these additional modes significantly affect the outcome only when the particle is notably slender. For most practical cases, the two-mode prolate squirmer model \citep{Theers2016,Qi2020,Chi2022,Vangogh2022,Gong2024} suffices to capture swimming dynamics. Thus, identifying the coefficients $B_1$ and $B_2$ completely determines the swimming mechanism here.

It is common to define the squirming ratio $\beta = B_2/B_1$ and in which case the sign of $\beta$ determines whether the propulsion of the swimmer originates from its front or rear. Organisms like \textit{Escherichia coli}, which generate propulsion from their rear, are classified as pushers and are characterized by $\beta < 0$. Conversely, organisms that draw fluid towards them from the front, such as \textit{Chlamydomonas reinhardtii}, are known as pullers and are characterized by $\beta > 0$. There are also swimmers whose propulsion mechanisms are not predominantly directed from the front or the rear; one example is \textit{Volvox carteri}, which has flagella evenly distributed across its spherical surface. For these organisms, referred to as neutral swimmers, $\beta = 0$.

\subsection{Newtonian fluid with viscosity gradients}
We endeavor to find the power expended by an active particle swimming in a Newtonian fluid with a spatially varying viscosity and thereafter determine the efficiency of that motion. The active particle is modeled as a prolate spheroid. The spheroid is an axisymmetric ellipsoid with two equal-length equatorial semi-axes $b$ and one longer polar semi-axis $a$ ($b \leq a$). The eccentricity
\begin{align}
e = \sqrt{1 - \frac{b^2}{a^2}}
\end{align}
represents the slenderness of the particle. When $e = 0$ ($b = a$), the particle is a sphere with radius $a$. The swimming direction $\boldsymbol{p}$ of the particle is the direction of its major axis. We consider a linearly varying viscosity field, $\eta(\boldsymbol{x})$, varying over a large macroscopic length scale $L$ such that
\begin{align}
\boldsymbol{\nabla} \eta = \frac{\eta_{\infty}}{L} \boldsymbol{d},
\end{align}
where $\eta_{\infty}$ is the viscosity at an arbitrary point near the particle, and  $\boldsymbol{d}$ is the unit direction of the viscosity gradient. Defining the dimensionless parameter, $\varepsilon = a/L \ll 1$, that characterizes the size of the particle versus the length scale of the gradient, we can write $\boldsymbol{\nabla} \eta = \varepsilon \frac{\eta_{\infty}}{a} \boldsymbol{d}$.

The flow around the particle in a Newtonian fluid with a spatially varying viscosity at low Reynolds number is governed by the continuity equation and Cauchy's equation of motion,
\begin{eqnarray}
\boldsymbol{\nabla} \boldsymbol{\cdot} \boldsymbol{u} & = & 0, \label{eqn:Stokes velocity}\\
\boldsymbol{\nabla} \boldsymbol{\cdot} \boldsymbol{\sigma} & = & \boldsymbol{0}, \label{eqn:Stokes stress}
\end{eqnarray}
where $\boldsymbol{u}$ is the velocity field and $\boldsymbol{\sigma} = - p \mathsf{\t I} + \eta(\boldsymbol{x}) \dot{\boldsymbol{\gamma}}$ is the stress tensor. Here $p$ is the pressure, $\dot{\boldsymbol{\gamma}} = \boldsymbol{\nabla} \boldsymbol{u} + (\boldsymbol{\nabla} \boldsymbol{u})^{\top}$ is strain-rate tensor. Defining an extra stress
\begin{align}
\boldsymbol{\tau}_{NN}   = (\eta (\boldsymbol{x}) - \eta_{\infty}) \dot{\boldsymbol{\gamma}},
\end{align}
we can alternatively write the stress tensor
 \begin{align}
\boldsymbol{\sigma}  = - p \mathsf{\t I} + \eta_{\infty} \dot{\boldsymbol{\gamma}} + \boldsymbol{\tau}_{NN}.
\end{align}

The boundary conditions on the flow field are that far away from the particle, the disturbance flow generated by the microswimmer should decay, which means
\begin{equation}
    \boldsymbol{u} \rightarrow \boldsymbol{0} \qquad \text{as} \, |\boldsymbol{r}| \rightarrow \infty.
\end{equation}
While on the particle surface (denoted by $S_p$), the fluid velocity $\boldsymbol{u}$ should satisfy no-slip conditions
\begin{equation}
    \boldsymbol{u} ( \boldsymbol{x} \in S_p) = \boldsymbol{U} + \boldsymbol{\Omega} \times \boldsymbol{r} + \boldsymbol{u}^s.
    \label{eq:bc}
\end{equation}
The slip velocity $\boldsymbol{u}^s$ is prescribed by \eqref{eqn:spheroidal_us}, while the translational and rotational velocities, $\boldsymbol{U}$ and $\boldsymbol{\Omega}$ are determined by satisfying the additional constraints that since the microswimmers are considered as neutrally buoyant there is no external force acting on the particle, and in the absence of particle inertia, the hydrodynamic force and torque on the particle must both vanish
\begin{eqnarray}
    \boldsymbol{F} = \int_{S_p} \boldsymbol{n} \boldsymbol{\cdot} \boldsymbol{\sigma} \, \text{d} S = \boldsymbol{0}, \label{eqn:BC3_a_f}\\
    \boldsymbol{L} = \int_{S_p} \boldsymbol{r} \times (\boldsymbol{n} \boldsymbol{\cdot} \boldsymbol{\sigma}) \, \text{d} S = \boldsymbol{0}. \label{eqn:BC3_a_t}
\end{eqnarray}

The power $\mathcal{P}$ expended by swimmer in its motion through the fluid is written as
\begin{align}
    \mathcal{P} &= -\int_{S_p} \boldsymbol{n} \boldsymbol{\cdot} \boldsymbol{\sigma} \boldsymbol{\cdot} \boldsymbol{u}  \, \text{d} S = -\int_{S_p} \boldsymbol{n} \boldsymbol{\cdot} \boldsymbol{\sigma} \boldsymbol{\cdot} \boldsymbol{u}^s  \, \text{d} S,
    \label{eqn:PowerS}
\end{align}
where the latter form comes from substitution of the boundary conditions \eqref{eq:bc} and application of the force and torque free conditions. Alternatively, an application of the divergence theorem yields the power expended by the swimmer in terms of the viscous dissipation in the fluid volume $\mathcal{V}$ in which the swimmer is immersed
\begin{align}
    \mathcal{P} &= \int_{\mathcal{V}} \boldsymbol{\sigma}:\boldsymbol{\nabla} \boldsymbol{u}  \, \text{d} V = \frac{1}{2}\int_{\mathcal{V}} \eta(\boldsymbol{x}) \dot{\boldsymbol{\gamma}}:\dot{\boldsymbol{\gamma}}  \, \text{d} V.
    \label{eqn:PowerV}
\end{align}

\citet{Lighthill1952}, defined the efficiency, $\mathcal{E}$, of a particle swimming at low Reynolds number as the ratio of the power, $\hat{\mathcal{P}}$, required to simply drag the particle (as a rigid body) at the same velocity as it swims, to the power expended by swimming $\mathcal{P}$, in other words
\begin{equation}
    \mathcal{E} = \frac{\hat{\mathcal{P}}}{\mathcal{P}}.
    \label{eqn:Efficiency}
\end{equation}

The towing power $\hat{\mathcal{P}}$ in \eqref{eqn:Efficiency} is given by
\begin{equation}
    \hat{\mathcal{P}} = -\hat{\boldsymbol{F}} \boldsymbol{\cdot} \boldsymbol{U} - \hat{\boldsymbol{L}} \boldsymbol{\cdot} \boldsymbol{\Omega},
\end{equation}
where the hydrodynamic force $\hat{\boldsymbol{F}}$ and torque $\hat{\boldsymbol{L}}$ due to undergoing the rigid body motion with velocities $\boldsymbol{U}$ and $\boldsymbol{\Omega}$ can be written
\begin{gather}
 \begin{pmatrix} \hat{\boldsymbol{F}}  \\ \hat{\boldsymbol{L}} \end{pmatrix}
 =-
 \begin{pmatrix}
   \mathsf{\t R}_{\boldsymbol{F U}} &
   \mathsf{\t R}_{\boldsymbol{F \Omega}} \\
   \mathsf{\t R}_{\boldsymbol{L U}} &
   \mathsf{\t R}_{\boldsymbol{L \Omega}}
   \end{pmatrix}\boldsymbol{\cdot}
   \begin{pmatrix}
    \boldsymbol{U} \\
    \boldsymbol{\Omega}
   \end{pmatrix},
   \label{eqn:resistance_matrix}
\end{gather}
where $(\mathsf{\t R}_{\boldsymbol{F U}},  \mathsf{\t R}_{\boldsymbol{F \Omega}}, \mathsf{\t R}_{\boldsymbol{L U}}, \mathsf{\t R}_{\boldsymbol{L \Omega}})$ are rigid body resistance tensors dependent on the eccentricity $e$, the orientation vector $\boldsymbol{p}$ and the viscosity gradient $\boldsymbol{\nabla} \eta$. The detailed expressions are given in the Appendix \ref{appendixb}.

\section{Asymptotic analysis}
Since the gradient in viscosity is considered weak, as characterized by $\varepsilon\ll 1$, we expand all flow terms $\{ \boldsymbol{u}, \boldsymbol{\sigma}, \boldsymbol{\tau}_{NN}, \dot{\boldsymbol{\gamma}}, \mathcal{P}, \hat{\mathcal{P}}, \mathcal{E} \}$ assuming a regular perturbation expansion in $\varepsilon$, for example $\boldsymbol{u} = \boldsymbol{u}_0 + \varepsilon \boldsymbol{u}_1 + O(\varepsilon^2)$.

The spatially varying viscosity field can be written in terms of the center of the particle $\boldsymbol{x}_c$
\begin{align}
\eta(\boldsymbol{x}) = \eta(\boldsymbol{x}_c) +(\boldsymbol{x}-\boldsymbol{x}_c) \boldsymbol{\cdot} \boldsymbol{\nabla} \eta = \eta(\boldsymbol{x}_c) +\varepsilon \frac{\eta_{\infty}}{a} \boldsymbol{d} \boldsymbol{\cdot} (\boldsymbol{x} - \boldsymbol{x}_c).
\end{align}
The center of the particle need not necessarily be in the plane $\mathcal{X}_\infty$ where $\eta(\boldsymbol{x}) = \eta_\infty$, but we can always define $\eta_{\infty}$ near the particle such that the distance $ | \boldsymbol{d} \boldsymbol{\cdot} (\boldsymbol{x} -\boldsymbol{x}_c) |/a \ll O(1/\varepsilon)$ for $\boldsymbol{x} \in \mathcal{X}_{\infty}$ and hence $\eta(\boldsymbol{x}_c) = \eta_\infty + O(\varepsilon)$.

\subsection{Homogeneous fluids}
At leading order, $\varepsilon\rightarrow 0$, the viscosity is homogeneous and so  we simply have an active particle swimming in a Newtonian fluid with a constant viscosity. The corresponding flow field is well known, for both a spherical and spheroidal active particles, and given in Appendix \ref{appendixa}. According to \eqref{eqn:PowerV}, the leading order value of power dissipation has the form,
\begin{equation}
        \mathcal{P}_0  = \frac{1}{2}\int_{\mathcal{V}} \eta_\infty \dot{\boldsymbol{\gamma}}_0:\dot{\boldsymbol{\gamma}}_0 \text{d} S.
\end{equation}
Evaluating the integral we obtain
\begin{equation}
        \mathcal{P}_0  = \pi a \eta_\infty (\mathcal{A} + \mathcal{B} \beta^2)B_1^2,
        \label{eqn: zeroth_order_power}
\end{equation}
where the terms $\mathcal{A}$ and $\mathcal{B}$ are functions only of the eccentricity and have the form,
\begin{align}
    \mathcal{A} & = \frac{2(1-e^2)[-2e + (1 + e^2) \mathcal{L}_e] }{e^3}, \\
    \mathcal{B} & = \frac{-8e^2(45 - 51e^2 + 8e^4) + 24e(15 - 22e^2 + 7e^4)\mathcal{L}_e + 6(-15 + 27e^2 - 13e^4 + e^6)\mathcal{L}_e^2}{3e^5[6e + (-3 + e^2)\mathcal{L}_e]},  \\
    \mathcal{L}_e & = \ln ((1+e)/(1-e)).
\end{align}

\citet{Keller1977} derived the above result for a neutral spheroidal squirmer, but here we have added the contribution of the second mode to obtain a formula valid for pushers and pullers. When $e \rightarrow 0$, $\mathcal{A}= 16/3$ and $\mathcal{B} = 8/3$, and we obtain the result for a spherical squirmer derived by \citet{Lighthill1952} and \citet{Blake1971}
\begin{equation}
    \mathcal{P}_{0,\text{sphere}} = \frac{8}{3} \pi (2  + \beta ^2) \eta_{\infty} a B_1^2 = 6 \pi a \eta_\infty (2  + \beta ^2)  U_{0,\text{sphere}}^2,
\end{equation}
where $U_{0,\text{sphere}} = 2B_1/3$ is the swimming speed of a spherical squirmer in Newtonian fluid with uniform viscosity. $\mathcal{A}$ and $\mathcal{B}$ are monotonically decreasing functions of eccentricity meaning power diminishes as a spheroidal squirmer becomes more slender and when $e \rightarrow 1$, $\mathcal{A} =\mathcal{B} = 0$, meaning that an infinitely slender squirmer requires no power to move (with speed equal to $B_1$) through a homogeneous Newtonian fluid.\\

The efficiency at leading order
\begin{equation}
    \mathcal{E}_0 = \frac{\hat{\mathcal{P}}_0}{\mathcal{P}_0}= \frac{\boldsymbol{U}_0 \boldsymbol{\cdot} \mathsf{\t R}_{\boldsymbol{F U}}^{(0)} \boldsymbol{\cdot}  \boldsymbol{U}_0}{\mathcal{P}_0}.
    \label{eqn:leading_order_SE}
\end{equation}
Using classic results of the rigid-body resistance (see Appendix \ref{appendixb}), and the swim speed of a spheroidal squirmer in a homogeneous Newtonian fluid \citep{Keller1977}
\begin{align}
\boldsymbol{U}_0 = \frac{2 e - (1 - e^2) \mathcal{L}_e}{2 e^3}B_1\boldsymbol{p},
\label{eqn:U0}
\end{align}
we obtain the efficiency
\begin{align}
\mathcal{E}_0 = \frac{\mathcal{F}}{\mathcal{A} + \mathcal{B} \beta^2},
\end{align}
where
\begin{align}
\mathcal{F} & = \frac{4[2e + (-1 + e^2) \mathcal{L}_e]^2}{e^3[-2e + (1 + e^2) \mathcal{L}_e]},
\end{align}
is monotonically decreasing function of slenderness $e$. When $e\rightarrow 0$, $\mathcal{F} = 8/3$ and we obtain the classic efficiency of a two-mode spherical squirmer
\begin{align}
\mathcal{E}_{0,\text{sphere}}=\frac{1}{2 + \beta^2}.
\end{align}
In the slender limit, $\mathcal{E}_0(e\rightarrow 1) = 3/\beta^2$ meaning efficiency is finite provided the presence of the wasteful $B_2$ mode. 

\subsection{Effect of gradients}
The effect of viscosity variations on the power dissipation in \eqref{eqn:PowerV} are captured at $O(\varepsilon)$ by 
\begin{equation}
    \varepsilon \mathcal{P}_1 = \frac{1}{2} \int_{\mathcal{V}} ( \eta(\boldsymbol{x}) - \eta_{\infty}) \dot{\boldsymbol{\gamma}}_0 : \dot{\boldsymbol{\gamma}}_0 \, \text{d} V.
    \label{eqn:first_order_power}
\end{equation}
Terms involving $\dot{\boldsymbol{\gamma}}_1$ are identically zero due to the force and torque free condition \citep{Nganguia2017, Corato2015} and so power at this order involves only integrating the leading order homogeneous flow field. With this formula, we obtain the power dissipation, valid up to $O(\varepsilon)$
\begin{align}
    \mathcal{P} =\pi (\mathcal{A} + \mathcal{B} \beta^2) \eta(\boldsymbol{x}_c) a B_1^2 +  \pi a^2 \mathcal{C} B_1^2 \beta( \boldsymbol{p} \boldsymbol{\cdot} \boldsymbol{\nabla} \eta),     \label{eqn:Power_diss_spheroid}
\end{align}
where
\begin{align}
\mathcal{C}  = \frac{4(-1 + e^2)[6e + (-3 + e^2) \mathcal{L}_e]}{e^5},
\end{align}
is a monotonically decreasing function of eccentricity. Taking the limit $e \rightarrow 0$, $\mathcal{C}= 32/15$ and we obtain for a sphere
\begin{equation}
    \mathcal{P}_{\text{sphere}} = \frac{8}{3} \pi (2 + \beta^2)  \eta(\boldsymbol{x}_c) a B_1^2 +   \frac{32}{15} \pi a^2 B_1^2\beta( \boldsymbol{p} \boldsymbol{\cdot} \boldsymbol{\nabla}\eta).  
    \label{eqn:Power_diss_spherical}
\end{equation}

What we see is that pushers ($\beta<0$) require less power to swim up viscosity gradients and more to swim down viscosity gradients in comparison to a homogeneous fluid, while the opposite is true for pullers ($\beta>0$). Neutral squirmers ($\beta=0$) do not see a change in power. The reason for this change in power expended can be understood in terms of the change in speed experienced by the squirmers. Previous research \citep{Datt2019,Gong2024} showed that pushers generate more thrust and thus swim faster and dissipate more energy when moving down viscosity gradients, conversely they generate less thrust, swim slower and dissipate less energy when moving up viscosity gradients. The opposite is true for pullers.

A spheroidal squirmer always dissipates less power (for the same $B_1$ and $B_2$ values), as $\mathcal{A}$, $\mathcal{B}$ and $\mathcal{C}$ are all monotonically decreasing functions of the slenderness $e$, that all go to zero when $e \rightarrow 1$. 

The correction to the efficiency, at $O(\varepsilon)$, can be written
\begin{align}
\mathcal{E}_1 = \frac{\hat{\mathcal{P}}_1}{\mathcal{P}_0}-\mathcal{E}_0\frac{\mathcal{P}_1}{\mathcal{P}_0}.
\end{align}
From Appendix \ref{appendixb}, we know that $\mathsf{\t R}_{\boldsymbol{F U}}$, $\mathsf{\t R}_{\boldsymbol{L \Omega}}$ are $O(1)$ at the leading order while $\mathsf{\t R}_{\boldsymbol{F \Omega}}, \mathsf{\t R}_{\boldsymbol{L U}}$ are $O(\varepsilon)$ as a result
\begin{align}
\hat{\mathcal{P}}_1 = 2 \boldsymbol{U}_1 \boldsymbol{\cdot} \mathsf{\t R}_{\boldsymbol{F U}}^{(0)} \boldsymbol{\cdot}  \boldsymbol{U}_0 + \boldsymbol{U}_0 \boldsymbol{\cdot} \mathsf{\t R}_{\boldsymbol{F U}}^{(1)} \boldsymbol{\cdot}  \boldsymbol{U}_0.
\end{align}
Using this result and the power obtained above we calculate the efficiency valid to $O(\varepsilon)$
\begin{align}
    \mathcal{E} = \frac{\mathcal{F}}{\mathcal{A} + \mathcal{B} \beta^2} + \varepsilon \frac{\mathcal{G}  + \mathcal{H} \beta^2}{(\mathcal{A} + \mathcal{B} \beta^2)^2} \beta(\boldsymbol{p} \boldsymbol{\cdot} \boldsymbol{d}),\label{eqn:Spheroidal_SE}
    \end{align}
where $\mathcal{G}$ (for $e < 0.964$) and $\mathcal{H}$ are monotonically decreasing function of slenderness
\begin{align}
    \mathcal{G} & = 8(1-e^2)[2e + (-1 + e^2) \mathcal{L}_e][8e^3(45+3e^2+2e^4)-4e^2(135 - 84e^2 + 5e^4 +4e^6)\mathcal{L}_e  \nonumber \\
    & \qquad + 2e(135 -177e^2 + 51e^4 + 5e^6 + 2e^8)\mathcal{L}_e^2 - 3(1-e^2)^2(15+e^4)\mathcal{L}_e^3] \nonumber \\
    & \qquad \times \{3e^8 [6e + (-3 + e^2) \mathcal{L}_e][-2e + (1 + e^2) \mathcal{L}_e] \}^{-1}, \\
    \mathcal{H} & = 4[2e + (-1 + e^2) \mathcal{L}_e]  \nonumber \\
    & \qquad [8e^2(45 - 51e^2+8e^4) - 24e(15 - 22e^2 + 7e^4)\mathcal{L}_e - 6(-15 +27e^2 - 13e^4 + e^6)\mathcal{L}_e^2] \nonumber \\
    & \qquad[4e^2(-9 + 3e^2 + 2e^4) - 4e(-9 + 6e^2 + e^6)\mathcal{L}_e + 3(-3 +3e^2 - e^4 + e^6)\mathcal{L}_e^2] \nonumber \\
    & \qquad \times \{9e^{10} [6e + (-3 + e^2) \mathcal{L}_e]^2 [-2e + (1 + e^2) \mathcal{L}_e] \}^{-1}.
\end{align}
For spheres $e\rightarrow 0$ the efficiency simplifies to 
\begin{equation}
    \mathcal{E}_{\text{sphere}} = \frac{1}{2 + \beta^2} + \varepsilon \frac{8 + 6 \beta^2}{5(2+ \beta^2)^2}\beta(\boldsymbol{p} \boldsymbol{\cdot} \boldsymbol{d}).\label{eqn:Spherical_SE}
\end{equation}

The essential message of these formulas is that, similar to speed and power, the correction to efficiency due to viscosity gradients is proportional to $\beta (\boldsymbol{p}\boldsymbol{\cdot}\boldsymbol{d})$, meaning that, for example, when swimming down viscosity gradients, pushers are faster, expend more power, but are more efficient, while pullers are slower, expend less power, and are less efficient. The opposite is true when swimming up viscosity gradients, however, we know that swimming up viscosity gradients is dynamically unstable and all squirmers tend to perform negative viscotaxis.

In deriving these formulas, we have assumed that the background viscosity remains fixed even in the presence of the particle. However, variations in viscosity generally arise from variations in an underlying field that affects the viscosity, such as temperature, salt or nutrient concentration and taking into account the effect of boundary conditions on the surface of the particle in relation to the underlying field will then lead to changes in the viscosity. In Appendix \ref{appendixc} we determine the effects of imposing a `no-flux' condition for viscosity at the surface of the particle and the associated effects on power expended and efficiency. Although there are quantitative differences in the parameters, the qualitative picture described above remains unchanged.

\section{Conclusion}
In this work we derived analytical formulas for the power expended by spheroidal squirmers swimming in linearly varying viscosity fields and the associated efficiency of that motion. We found that pushers are faster and more efficient when moving down gradients but slower and less efficient moving up viscosity gradients, and the opposite is true for pullers. We also evaluated the effect of shape on power expenditure and efficiency when swimming in viscosity gradients and found that the change in both due to gradients monotonically diminishes with increasing slenderness.

Swimming down viscosity gradients is favourable for pushers because they generate thrust from the rear, where the viscosity is highest, this leads to faster swimming and thus a greater level of power expenditure; however, the energetic cost of this boost is lower (relative to rigid body motion) in comparison to a homogeneous fluid thus the increase in efficiency. While both pushers and pullers can minimize or maximize efficiency depending on orientation relative to the gradient, both display negative viscotaxis, which means that pushers dynamically tend to the most efficient orientation while pullers the least. However, this conclusion is sensitive to geometry. For example, if we construct a pusher made from a very thin `tail' that generates thrust and a large head that bears the majority of the drag, then this sort of swimmer would display \textit{positive} viscotaxis \citep{Gong2024}, and be efficiency minimizing.  This shows how shape and gait play a critical role in driving dynamics in inhomogeneous environments and how both efficiency minimizing and maximizing dynamical states can be stable depending on the specific shape and gait. This raises questions about whether organisms might adjust their swimming gait in light of this fact. Finally we note, that while a puller is less efficient moving down a viscosity gradient in comparison to a homogeneous fluid of the same viscosity, it will still be energetically preferable to move to the region of lower viscosity, even if the instantaneous dynamics are slower and less efficient.\\


\noindent\textbf{Funding.} This work was supported by the Natural Sciences and Engineering Research Council of Canada (RGPIN-2020-04850) and by a UBC Killam Accelerator Research Fellowship to G.J.E.\\

\noindent\textbf{Declaration of interests.} The authors report no conflict of interest.

\appendix
\section{An active prolate spheroid in Stokes flow}\label{appendixa}
The flow field $\boldsymbol{u}_0$ of an active spheroid swimming in a Newtonian fluid with uniform viscosity can be expressed in terms of a stream function $\psi_0$ \citep{Keller1977,Theers2016,Vangogh2022},
  \begin{equation}
      \boldsymbol{u}_0 =  \frac{1}{c^2 \sqrt{\zeta_1^2 - \zeta_2^2}} (\frac{1}{\sqrt{\zeta_1^2 - 1}}\frac{\partial \psi_0}{\partial \zeta_2} \boldsymbol{e}_{\zeta_1} -  \frac{1}{ \sqrt{1 - \zeta_2^2 }} \frac{\partial \psi_0}{\partial \zeta_1} \boldsymbol{e}_{\zeta_2}),
      \label{active_spheroid_flow_field}
  \end{equation}
using prolate spheroid coordinates $O'\zeta_1 \zeta_2 \phi$, where $O'$ is the center of the spheroid at $\boldsymbol{x}_c$. The spheroidal coordinate system can be related to a particle-aligned Cartesian coordinate system $O'X_1X_2X_3$ as follows
\begin{align}
    X_1 & = c \sqrt{\zeta_1^2 - 1} \sqrt{ 1 - \zeta_2^2} \cos \phi, \nonumber \\
    X_2 & = c \sqrt{\zeta_1^2 - 1} \sqrt{ 1 - \zeta_2^2} \sin \phi,  \\
    X_3 & = c \zeta_1 \zeta_2, \nonumber
\end{align}
where $ 1 \leqslant \zeta_1 < \infty$, $ -1 \leqslant \zeta_2 \leqslant 1$ and $0 \leqslant \phi < 2\pi $. $c = \sqrt{a^2 - b^2}$ is half of the focal length. We can define $ \tilde{\zeta_1} = 1 / e$ with $\zeta_1 > \tilde{\zeta}_1$ corresponding to the fluid domain exterior to the surface $\zeta_1 = \tilde{\zeta}_1$ of the particle.
  
The stream function $\psi_0$ has the solution
 \begin{align}
     \psi_0 & = C_1 H_2 (\zeta_1) G_2 (\zeta_2) + C_2 \zeta_1 ( 1 - \zeta_2^2) \nonumber \\
     & \quad + C_3 H_3 (\zeta_1) G_3 (\zeta_2) + C_4 \zeta_2 ( 1 - \zeta_2^2) + \frac{1}{2} U_0 c^2 (\zeta_1^2 - 1)(1 - \zeta_2^2).
 \end{align}
 Here $H_n(x)$ and $G_n(x)$ are Gegenbauer functions of the first and second order of degree $-1/2$ \citep{Theers2016}. The coefficients $C_n$ are
 \begin{align}
     C_1 & = 2c^2 \frac{U_0 (\tilde{\zeta}_1^2 + 1) - 2 B_1 \tilde{\zeta}_1^2}{- \tilde{\zeta}_1 + (1 + \tilde{\zeta}_1^2 ) \coth^{-1} \tilde{\zeta}_1}, \nonumber \\
     C_2 & = c^2 \frac{B_1  \tilde{\zeta}_1 [\tilde{\zeta}_1 - (\tilde{\zeta}_1^2 - 1) \coth^{-1} \tilde{\zeta}_1) - U_0]}{- \tilde{\zeta}_1 + (1 + \tilde{\zeta}_1^2) \coth^{-1} \tilde{\zeta}_1}, \nonumber \\
     C_3 & = c^2 \frac{ 4 B_2 \tilde{\zeta}_1}{3 \tilde{\zeta}_1 + (1 - 3\tilde{\zeta}_1^2) \coth^{-1} \tilde{\zeta}_1}, \nonumber \\
     C_4 & = c^2 \frac{ B_2 \tilde{\zeta}_1 [2/3 - \tilde{\zeta}_1^2 + \tilde{\zeta}_1(\tilde{\zeta}_1^2 - 1) \coth^{-1} \tilde{\zeta}_1]}{3 \tilde{\zeta}_1 + (1 - 3\tilde{\zeta}_1^2) \coth^{-1} \tilde{\zeta}_1},
 \end{align}
 and $U_0 = \left|\boldsymbol{U}_0\right|$ given in \eqref{eqn:U0}.

\section{Resistance tensor for a passive prolate spheroid in viscosity gradient}\label{appendixb}
Here we derive the resistance tensor for a spheroidal particle undergoing rigid-body motion in a fluid with a linearly varying viscosity field. We follow similar steps to those shown in a previous paper deriving the mobility of a spheroidal particle in a viscosity gradient \citep{Gong2024} and likewise we use compact six dimensional vectors for velocities $\mathsf{\t U} = (\boldsymbol{U}, \boldsymbol{\Omega})^{\top}$ and forces and torques $\hat{\mathsf{\t F}} = ( \hat{\boldsymbol{F}}, \hat{\boldsymbol{L}})^{\top}$ to simplify formulas. Following the notation of this paper, the hat notation corresponds to values associated with rigid-body motion.

The resistance tensor to leading order, $\mathsf{\t R}^{(0)}_{\mathsf{\t F} \mathsf{\t U}}$, corresponding to a homogeneous fluid with uniform viscosity $\eta_\infty$ is well known and given by \citet{Kim1991}. The hydrodynamic force and torque in a viscosity gradient can be written as
\begin{equation}
    \hat{\mathsf{\t F}} =  -\mathsf{\t R}^{(0)}_{\mathsf{\t F} \mathsf{\t U}} \boldsymbol{\cdot} \mathsf{\t U} + \hat{\mathsf{\t F}}_{NN}.
\end{equation}
The `extra' hydrodynamic force and torque corresponding to the extra stress $\hat{\boldsymbol{\tau}}_{NN}$ from the gradient can be shown, by the reciprocal theorem, to be
\begin{equation}
    \hat{\mathsf{\t F}}_{NN} = - \int_{\mathcal{V}} \hat{\boldsymbol{\tau}}_{NN} : \hat{\mathsf{\t E}}_{\mathsf{\t U}}^{(0)} \, \text{d} V, 
\end{equation}
where the operator $\hat{\mathsf{\t E}}_{\mathsf{\t U}}^{(0)}$ corresponds to the rate of strain tensor in a homogeneous fluid,
\begin{equation}
    \hat{\dot{\boldsymbol{\gamma}}}_0 = 2 \hat{\mathsf{\t E}}_{\mathsf{\t U}}^{(0)} \boldsymbol{\cdot} \mathsf{\t U}, \label{eqn:passive_strain}
\end{equation}
and may be calculated with the flow field due to a spheroid undergoing rigid-body motion in a homogeneous Newtonian fluid, given by \citet{Kim1991}. Substitution of $\hat{\boldsymbol{\tau}}_{NN} = (\eta (\boldsymbol{x}) - \eta_{\infty}) \hat{\dot{\boldsymbol{\gamma}}}$ into the equation above, and expanding terms in powers of $\varepsilon$ we obtain
\begin{equation}
    \hat{\mathsf{\t F}}_{NN} = -\int_{\mathcal{V}} 2 (\eta (\boldsymbol{x}) - \eta_{\infty}) \hat{\mathsf{\t E}}_{\mathsf{\t U}}^{(0)} : \hat{\mathsf{\t E}}_{\mathsf{\t U}}^{(0)}\boldsymbol{\cdot} \mathsf{\t U}\, \text{d} V  + O (\varepsilon^2), \label{eqn:F_NN}
\end{equation}
hence we can identify
\begin{equation}
\varepsilon \mathsf{\t R}^{(1)}_{\mathsf{\t F} \mathsf{\t U}} = \int_{\mathcal{V}} 2 (\eta (\boldsymbol{x}) - \eta_{\infty}) \hat{\mathsf{\t E}}_{\mathsf{\t U}}^{(0)} : \hat{\mathsf{\t E}}_{\mathsf{\t U}}^{(0)}\, \text{d} V.
\end{equation}

Combining terms, the resistance tensor accurate to $O(\varepsilon)$ is $\mathsf{\t R}_{\mathsf{\t FU}} = \mathsf{\t R}^{(0)}_{\mathsf{\t F} \mathsf{\t U}} + \varepsilon\mathsf{\t R}^{(1)}_{\mathsf{\t F} \mathsf{\t U}}$, where
\begin{gather}
 \boldsymbol{\mathsf{\t R}}_{\mathsf{\t FU}} = 
 \begin{pmatrix}
   \mathsf{\t R}_{\boldsymbol{F} \boldsymbol{U}} 
   \quad \mathsf{\t R}_{\boldsymbol{F} \boldsymbol{\Omega}} \\
   \mathsf{\t R}_{\boldsymbol{L} \boldsymbol{U}} 
   \quad \mathsf{\t R}_{\boldsymbol{L} \boldsymbol{\Omega}}
   \end{pmatrix},
   \label{eqn:resistance_matrix_2}
\end{gather}
and $\mathsf{\t R}_{\boldsymbol{F} \boldsymbol{\Omega}} = \mathsf{\t R}_{\boldsymbol{L} \boldsymbol{U}}^{\top}$. The resistance depends on the eccentricity $e$, the orientation vector $\boldsymbol{p}$ and the viscosity gradient $\boldsymbol{\nabla} \eta$. Detailed expressions for all couplings in \eqref{eqn:resistance_matrix_2} are,
\begin{align}
    \mathsf{\t R}_{\boldsymbol{FU}} & = 6 \pi \eta (\boldsymbol{x}_c) a [ \mathcal{X}^{A} \boldsymbol{p} \boldsymbol{p} + \mathcal{Y}^{A} (\mathsf{\t I} - \boldsymbol{p} \boldsymbol{p})], \\
    \mathsf{\t R}_{\boldsymbol{L\Omega}} & = 8 \pi \eta (\boldsymbol{x}_c) a^3 [ \mathcal{X}^{C} \boldsymbol{p} \boldsymbol{p} + \mathcal{Y}^{C} (\mathsf{\t I} - \boldsymbol{p} \boldsymbol{p})], \\
    \mathsf{\t R}_{\boldsymbol{F\Omega}} = \mathsf{\t R}_{\boldsymbol{L U}}^{\top} & = \varepsilon6 \pi \eta_{\infty} a^2  \Big [\lambda_1 ( \boldsymbol{d} \times \mathsf{\t I}) + \lambda_2 ( \boldsymbol{p} \boldsymbol{\cdot} \boldsymbol{d}) ( \boldsymbol{p} \times \mathsf{\t I}) + \lambda_3 \boldsymbol{p} (\boldsymbol{d} \times \boldsymbol{p}) \Big ],\label{eqn:B}
\end{align}
where
\begin{gather}\allowdisplaybreaks
    \lambda_1  = \frac{16 e^3 (1-e^2)}{9[ - 2e + (1 - 3 e^2) \mathcal{L}_e]},   \\
    \lambda_2  = \frac{8[ 6e^4(1 - e^2) + e^3 ( -3 + 4e^2 + 3e^4) \mathcal{L}_e]}{9 [-2e + (1 + e^2) \mathcal{L}_e][ - 2e + (1 - 3 e^2) \mathcal{L}_e]},  \\
    \lambda_3  = \frac{4[e^5(36 - 28e^2) + e^4(-36 + 40e^2 + 4e^4)\mathcal{L}_e + e^3(9 - 13e^2 - e^4 + 5e^6)\mathcal{L}_e^2]}{9 [-2e + (1 + e^2) \mathcal{L}_e]^2[ - 2e + (1 - 3 e^2) \mathcal{L}_e]},\\
        \mathcal{X}^A =\frac{8 e^3}{3[ - 2e +  \left(1 + e^2\right)\mathcal{L}_e]},\\
    \mathcal{Y}^A =\frac{16 e^3}{3[ 2e +  \left(3 e^2 - 1\right)\mathcal{L}_e]},\\
    \mathcal{X}^C =\frac{4e^3(1 - e^2)}{3 [ 2e -  \left( 1 - e^2\right)\mathcal{L}_e  ]},\\
    \mathcal{Y}^C =\frac{4e^3(2 - e^2)}{ 3 [-2e +  \left( 1 + e^2\right)\mathcal{L}_e  ]},
\end{gather}
are functions of eccentricity $e$.

\section{Disturbance viscosity effects}\label{appendixc}
Variations in viscosity generally arise from variations in an underlying field that affects the viscosity, such as temperature, salt or nutrient concentration. Taking into account the effect of boundary conditions on the surface of the particle in relation to the underlying field will then lead to changes in the viscosity. For example, in an otherwise linear salt concentration field, the presence of a particle will disrupt the field due to salt impermeability. Although these disturbances diminish with distance from the particle, the disturbance does have a leading-order effect on the dynamics of the active particle \citep{Shaik2021}.

Here we determine the power dissipation and swimming efficiency, factoring in the disturbance to viscosity due to the application of a no-flux condition at the particle's surface. The total viscosity field can be represented as the superposition of an ambient viscosity field (denoted as $\eta_0$) and a disturbance viscosity field (denoted by prime),
\begin{equation}
	\eta = \eta_0 + \eta'.
\end{equation}

The boundary conditions on the disturbance viscosity field $\eta'$ are as follows. The disturbance viscosity should diminish in the far-field region,
\begin{equation}
        \eta' \rightarrow 0 \qquad \text{as} \thickspace |\boldsymbol{r}| \rightarrow \infty.
\end{equation}

The transport of scalar fields, such as temperature, salt, or nutrient concentrations, can be characterized using an advection-diffusion equation. When variations in these scalar fields are minimal, changes in viscosity can be modeled with a similar advection-diffusion equation. For slow-moving microswimmers in environments with highly diffusive scalar fields, such as temperature or salt concentration, advection is minimal, resulting in the viscosity distribution following the Laplace equation. Since the ambient viscosity field is linear, the disturbance viscosity field must also conform to the Laplace equation,
\begin{equation}
     \nabla^2 \eta = \nabla^2 \eta' = 0.
\end{equation}

The disturbance viscosity field is also governed by the boundary conditions on the particle's surface. We consider the surface to be impermeable to salt or insulated against temperature changes and this leads to an associated `no-flux' condition for viscosity at the particle's boundary,
\begin{equation}
    \boldsymbol{n} \boldsymbol{\cdot}\boldsymbol{\nabla} \eta = 0, \qquad \text{on} \medspace \medspace S_p.
\end{equation}

Supposing the ambient viscosity field is along $\boldsymbol{e}_1$, the mathematical expression of the disturbance viscosity field $\eta'$ in a particle-aligned coordinate is,
\begin{equation} 
    \eta ' = A_{1,0}  P_1^0 (\zeta_2) Q_1^0 (\zeta_1) + A_{1,1}  P_1^1 (\zeta_2) Q_1^1 (\zeta_1) \cos (\phi)
\end{equation}
where
\begin{align}
     A_{1,0} & = \varepsilon \eta_{\infty} p_1 \frac{2e (1 - e^2)}{ \left[ 2e - \left(1 - e^2\right) \mathcal{L}_e\right]}, \nonumber \\
     A_{1,1} & = \varepsilon \eta_{\infty} \sqrt{1 - p_1^2} \frac{2e (1 - e^2)}{ \left[ 2e - 4 e^3 - \left(1 - e^2\right) \mathcal{L}_e\right]}. \nonumber 
\end{align}
Here $p_1 = \boldsymbol{p} \boldsymbol{\cdot} \boldsymbol{e}_1$. $P_k^m$ and $Q_k^m$ are denoted as the associated Legendre polynomial of the first and second kind, with the degree $k$ and the order $m$. Their detailed mathematical formulations can be referenced in \citet{Abramowitz1964}.

The power dissipation of a prolate spheroidal squirmer with a no-flux condition is, to first order,
\begin{equation}
    \mathcal{P} =\pi (\mathcal{A} + \mathcal{B} \beta^2) \eta(\boldsymbol{x}_c) a B_1^2 +  \pi a^2 \mathcal{C}^{nf}  B_1^2 \beta( \boldsymbol{p} \boldsymbol{\cdot} \boldsymbol{\nabla} \eta),
\end{equation}
where,
\begin{equation}
	 \mathcal{C}^{nf} = \frac{8(-1 + e^2)[6e + (-3 + e^2) \mathcal{L}_e]}{e^2[2e + (-1 + e^2) \mathcal{L}_e]},
\end{equation}
is a monotonically decreasing function of eccentricity. For a sphere, $e \rightarrow 0$, $\mathcal{C}^{nf} = 16/5$, and we have,
\begin{equation}
    \mathcal{P}_{\text{sphere}}^{nf} = \frac{8}{3} \pi (2 + \beta^2)  \eta(\boldsymbol{x}_c) a B_1^2 +   \frac{16}{5} \pi a^2 B_1^2\beta( \boldsymbol{p} \boldsymbol{\cdot} \boldsymbol{\nabla}\eta). 
\end{equation}
The corresponding swimming efficiency up to $O(\varepsilon)$ is,
\begin{equation}
\mathcal{E} = \frac{\mathcal{F}}{\mathcal{A} + \mathcal{B} \beta^2} + \varepsilon \frac{\mathcal{G}^{nf}  + \mathcal{H}^{nf}  \beta^2}{(\mathcal{A} + \mathcal{B} \beta^2)^2} \beta(\boldsymbol{p} \boldsymbol{\cdot} \boldsymbol{d}),
\end{equation}
where $\mathcal{G}^{nf}$ and $\mathcal{H}^{nf}$ follows,
\begin{align}
    \mathcal{G}^{nf} & = 8(1-e^2)[8e^3(45+87e^2-32e^4)+4e^2(-135 - 36e^2 + 19e^4 + 32e^6)\mathcal{L}_e  \nonumber \\
    & \qquad + (270e - 426e^3 + 402e^5 - 182e^7)\mathcal{L}_e^2 + (-45 + 186e^2 - 212e^4 + 54e^6 + 17e^8)\mathcal{L}_e^3 \nonumber \\
    & \qquad - 12e (1-e^2)^2 (1+e^2) \mathcal{L}_e^4 ]\times \{3e^5 [6e + (-3 + e^2) \mathcal{L}_e][-2e + (1 + e^2) \mathcal{L}_e] \}^{-1}, \\
    \mathcal{H}^{nf} & = 8 [-4e^2(45 - 51e^2+8e^4) + 12e(15 - 22e^2 + 7e^4)\mathcal{L}_e -3(15 -27e^2 + 13e^4 - e^6)\mathcal{L}_e^2] \nonumber \\
    & \qquad[4e^2(63 - 87e^2 + 32e^4) - 4e(63 - 96e^2 + 37e^4)\mathcal{L}_e + (63 - 81e^2 + e^4 + 17e^6)\mathcal{L}_e^2 \nonumber \\
    & \qquad -12e(1-e^2)^2 \mathcal{L}_e^3] \times \{9e^{7} [6e + (-3 + e^2) \mathcal{L}_e]^2 [-2e + (1 + e^2) \mathcal{L}_e] \}^{-1}.
\end{align}
Here $\mathcal{G}^{nf}$(for $e < 0.954$) and $\mathcal{H}^{nf}$ are monotonically decreasing functions of slenderness and they both vanish when $e \rightarrow 1$. For a sphere we obtain,
\begin{equation}
\mathcal{E}_{\text{sphere}}^{nf} = \frac{1}{2 + \beta^2} + \varepsilon \frac{14 + 13 \beta^2}{10(2+ \beta^2)^2}\beta(\boldsymbol{p} \boldsymbol{\cdot} \boldsymbol{d}).
\end{equation}
We see that taking into account the disturbance viscosity due to a no-flux boundary condition does not change the qualitative picture when compared to the effects of ambient viscosity alone.

\bibliography{efficiency}

\end{document}